\documentclass[prb,showpacs]{revtex4}
\usepackage{amssymb}
\usepackage{graphicx}
\usepackage{dcolumn}
\usepackage{bm}
\usepackage[normalem]{ulem}

\begin{document}

\title{Correlated bosons on a lattice: Dynamical mean-field theory for \\
Bose-Einstein condensed and normal phases }
\author{Krzysztof Byczuk$^{1,2}$ and Dieter Vollhardt$^1$ }
\affiliation{\centerline{$^1$ Theoretical Physics III, Center for Electronic
  Correlations and Magnetism,
Institute for Physics,}\\
\centerline{University of Augsburg, 86135 Augsburg, Germany, }\\
\centerline {$^2$ Institute of Theoretical Physics,
University of Warsaw, ul. Ho\.za 69, 00-681 Warszawa, Poland } }
\date{\today }

\begin{abstract}
We formulate a bosonic dynamical mean-field theory (B-DMFT) which provides a
comprehensive, thermodynamically consistent framework for the theoretical
investigation of correlated lattice bosons. The B-DMFT is applicable for
arbitrary values of the coupling parameters and temperature and becomes
exact in the limit of high spatial dimensions $d$ or coordination number $Z$
of the lattice. In contrast to its fermionic counterpart the construction of
the B-DMFT requires different scalings of the hopping amplitudes with $Z$
depending on whether the bosons are in their normal state or in the
Bose-Einstein condensate. A detailed discussion of how this conceptual
problem can be overcome by performing the scaling in the action rather than
in the Hamiltonian itself is presented. The B-DMFT treats normal and
condensed bosons on equal footing and thus includes the effects caused by
their dynamic coupling. It reproduces all previously investigated limits in
parameter space such as the Beliaev-Popov and Hartree-Fock-Bogoliubov
approximations and generalizes the existing mean-field theories of
interacting bosons. The self-consistency equations of the B-DMFT are those
of a bosonic single-impurity coupled to two reservoirs corresponding to
bosons in the condensate and in the normal state, respectively. We employ
the B-DMFT to solve a model of itinerant and localized, interacting bosons
analytically. The local correlations are found to \emph{enhance} the
condensate density and the Bose-Einstein condensate (BEC) transition
temperature $T_{\mathrm{BEC}}$. This effect may be used experimentally to
increase $T_{\mathrm{BEC}}$ of bosonic atoms in optical lattices.
\end{abstract}

\pacs{71.10.Fd, 67.85.Hj}
\maketitle

\section{Introduction}

The observation of Bose-Einstein condensation (BEC) in ultra-cold,
atomic gases has greatly stimulated research into the properties of
this fascinating quantum state of matter. \cite{bec} In particular,
experiments with alkali atoms confined in optical lattices
\cite{Greiner02,lewenstein06,Bloch07} have renewed the theoretical
interest \cite{jaksch98,Isacsson05,Huber07} in the physics of
strongly correlated bosons on lattices, which promises significant
new insights and even applications in fields such as quantum
computing. \cite{micheli06}

The investigation of correlated lattice bosons is not only relevant
for ultra-cold bosonic atoms in optical lattices, but has a long
history starting with Matsubara and Matsuda's
\cite{matsubara56,matsubara57,morita} formulation of a lattice model
of liquid $^4$He. A new direction of research was initiated by
Fisher \textit{et al}., \cite{fisher89} who studied lattice bosons
with and without disorder to explore the  superfluid-insulator
transition \cite{Bartouni90,rok,Sheshadri93,freericks94,freericks96}
and boson localization observed in $^4$He absorbed in porous media.
\cite{reppy92} Granular superconductors forming weak Josephson
junctions have also been described by interacting lattice bosons.
\cite{kampf93,bruder05} Recently quantum phase transitions in
magnetic systems such as TlCuCl$_3$, \cite{ruegg03} which can be
induced by tuning the magnetic field, have been interpreted as the
BEC of magnons.\cite{Giamarchi} Bosonic supersolids
\cite{Kim04,leggett01} promise to be yet another fascinating state
of bosonic matter.

In this paper we formulate the first comprehensive, thermodynamically
consistent theory of correlated lattice boson systems, namely a bosonic
dynamical mean-field theory (B-DMFT) which is applicable for arbitrary
values of the coupling parameters and temperature. The B-DMFT includes all
local, dynamical correlations of the many-boson system and becomes exact in
the limit of infinite space dimensions in analogy with its successful
fermionic counterpart. \cite{metzner89,georges96,pruschke95,kotliar04} With
the B-DMFT we are able to solve a lattice model of itinerant and localized,
interacting bosons. The local correlations lead to an enhancement both of
the BEC transition temperature $T_{BEC}$ and the condensate fraction as
compared to the non-interacting system. Hence bosonic correlations can be
employed in the laboratory to reach higher values of $T_{BEC}$.

This paper is organized as follows: In Sec. II we introduce the bosonic
Hubbard model and explain the specific problems arising in the construction
of a bosonic dynamic mean-field theory (B-DMFT) in the limit of large
coordination number $Z$ of the lattice, namely, the problem of how to scale
the hopping amplitudes with $Z$. The self-consistency equations and the
general structure of the B-DMFT are discussed in Sec. III. The comprehensive
nature of the B-DMFT is demonstrated in Sec. IV by explicitly reproducing
results previously obtained in special limits of parameter space and by
deriving other bosonic mean-field theories. In Sec. V the B-DMFT is employed
to solve a bosonic version of the Falicov-Kimball model and it is shown that
correlation effects lead to an enhancement of $T_{BEC}$. Conclusions and an
outlook in Sec. VI close the presentation.

\section{Correlated lattice bosons}

\subsection{Generalized bosonic Hubbard model}

In the following we consider a many-particle system with different
species of bosons as it can be realized in optical lattices.
\cite{lewenstein06,Bloch07} This may either involve different atoms
as, for example, in a binary mixture of $^{87}$Rb and $^7$Li, or one
type of atom in different hyper-fine quantum states such as
$^{87}$Rb, where the total nuclear spin $I=3/2$ adds to the spin
$S=1/2$ of the valence s-electron, giving states with $F=I+S=1$ or
$2$. Such systems may be modelled by a generalized bosonic Hubbard
Hamiltonian\cite{matsubara56,gersch63,fisher89,jaksch98}
\begin{equation}
H=\sum_{ij\nu}t_{ij}^{\nu}b^{\dagger}_{i\nu} b_{j\nu} +
\frac{1}{2}\sum_{i\mu\nu}U_{\mu\nu} n_{i\mu}
(n_{i\nu}-\delta_{\mu\nu})\equiv H_0+H_{\mathrm{int}},
\label{hamiltonian}
\end{equation}
where $n_{i\nu}=b^{\dagger}_{i\nu} b_{i\nu}$ is the occupation number
operator for bosons of species $\nu$. Furthermore, $t_{ij}^{\nu}$ are
hopping amplitudes of $\nu$-bosons and $U_{\mu\nu}$ are local
density-density interactions between $\mu$- and $\nu$-bosons on the same
lattice site. An exchange interaction for spinor bosons can be easily
included. In general the many-boson model (\ref{hamiltonian}) in unsolvable.

\subsection{Construction of a comprehensive mean-field theory}

The explanation of experiments with correlated lattice bosons in
quantum optics and condensed matter physics requires a comprehensive
theoretical scheme for the investigation of the Hamiltonian
(\ref{hamiltonian}). In particular, it must be capable of describing
thermal and quantum phase transitions and thus provide the phase
diagram and the thermodynamics for the entire range of microscopic
parameters. In the case of lattice fermions such a framework already
exists: the dynamical mean-field theory (DMFT) \cite{georges96}.
Indeed, the DMFT has proved to be a very successful, comprehensive
mean-field theory for models and materials with strongly correlated
electrons. \cite{kotliar04,kotliar06} In particular, it provides a
quantitative description of the Mott-Hubbard metal-insulator
transition, photoemission spectra, magnetic phases, and other
correlation induced phenomena. The DMFT has the virtue of becoming
exact in the limit of infinite space dimensions $d$ or,
equivalently, infinite coordination number $Z$, i.e., number of
nearest neighbors ($Z=2d$ for a $d$-dimensional hypercubic lattice).
\cite{metzner89} This limit is well-known to produce mean-field
theories  which are diagrammatically controlled and whose free
energies have no unphysical singularities (e.g., the Weiss
mean-field theory for the Ising or Heisenberg spin
models).\cite{Itzykson} To obtain a physically meaningful mean-field
theory the free energy of the model has to remain finite in the
limit $d$ or $Z\rightarrow \infty $. \cite{metzner89} This requires
a suitable scaling of the coupling parameters with $d$ or $Z$, e.g.,
$J\rightarrow {\tilde{J}}/Z$, ${\tilde{ J}}$= const., for Ising
spins with nearest-neighbor coupling $J$. While for the Ising model
the scaling is self-evident this is not so for more complicated
models. Namely, fermionic or bosonic many-particle systems are
usually described by a Hamiltonian consisting of several
non-commuting terms each of which is associated with a coupling
parameter, e.g., a hopping amplitude or interaction. In such a case
the question of how to scale these parameters has no unique answer
since this depends on the physical effects one wishes to
explore.\cite{mh89,si96} In any case, the scaling should be
performed such that the model remains non-trivial and its free
energy stays finite in the $Z\rightarrow \infty $ limit. By
``non-trivial'' we mean that not only $\langle H_{0}\rangle $ and
$\langle H_{\mathrm{int}}\rangle $, but also the \emph{competition}
between these terms as expressed by $\langle \lbrack
H_{0},H_{\mathrm{int}}]\rangle $, should remain finite; here
$\langle ...\rangle $ denotes the quantum and statistical average of
operators.  In the literature on lattice bosons the $d\rightarrow
\infty $ limit was so far considered only in connection with the
distance-independent (``infinite-range'') hopping of the bosons
\cite{rok,freericks94,freericks96} in which the mean-field theory of
Fisher \textit{et al}. \cite{fisher89} for the Bose-condensed phase
becomes exact. As will be discussed below this is a static
mean-field theory since normal and condensed bosons are not
dynamically coupled. In particular, in the normal phase one has
\textbf{$\langle \lbrack H_{0},H_{\mathrm{int}}]\rangle =0$} and the
lattice problem is reduced to a single-site (``atomic'') problem
where particles are immobile. Another static mean-field theory is
the Bogoliubov approximation, which yields a good weak-coupling
mean-field theory for bosons in a continuum. For lattice bosons this
approximation fails to describe the Mott superfluid-insulator
transition.

Evidently, a \textit{dynamical} mean-field theory is needed to
describe the rich physics of interacting lattice bosons, e.g. cold
atoms in optical lattices, \cite{Greiner02} within one conceptual
framework. A comprehensive DMFT for correlated lattice bosons, i.e.,
a theory which can describe normal \emph{and} condensed bosons on
the same footing, did not exist up to now. In the following we will
discuss the conceptual problems which prevented the formulation of
such a bosonic DMFT, and how they can be overcome.

\subsection{Lattice bosons in infinite dimensions: Different scaling for Bose-Einstein condensed and normal bosons}

A macroscopically large number of bosons can condense into a single
quantum state. This BEC may be detected in the spectral
decomposition of the one-particle density matrix
\begin{equation}
\rho _{ij}=\langle b_{i}^{\dagger }b_{j}\rangle =\sum_{\alpha }\lambda
_{\alpha }\phi _{i\alpha }^{\ast }\phi _{j\alpha },
\end{equation}
where $b_{i}^{\dagger }$ and $b_{i}$ are creation and annihilation
operators, respectively, for a boson at a lattice site $i$, with
$\phi _{i\alpha }$ as the corresponding wave function. For
simplicity we discuss here only a single species of bosons so the
index $\nu $ can be omitted. When BEC occurs one of the eigenvalues
becomes macroscopically large, $\lambda _{0}=N_{0}\sim O(N)$, where
$N$ is the total number of bosons. The density matrix then
decomposes into
\begin{equation}
\rho _{ij}=N_{0}\phi _{i0}^{\ast }\phi _{j0}+\tilde{\rho}_{ij},
\end{equation}%
where the second term corresponds to non-condensed, "normal" bosons. The
first term has the remarkable property that it does not decrease even at
large distance $R_{ij}=||\mathbf{R}_{i}-\mathbf{R}_{j}||$ between the bosons
at sites $i$ and $j$. Here $||\mathbf{R}||$ denotes the length of $\mathbf{R}
$ obtained by counting the minimal number of links between two sites on a
lattice. By contrast, the second term in $\rho _{ij}$ decreases with
increasing $R_{ij}$.

This has immediate consequences for the kinetic part of the Hamiltonian
\begin{equation}
H_0=-t\sum_{<ij>}b^{\dagger}_ib_j  \label{nonhamiltonian}
\end{equation}
with $-t$ as the amplitude for hopping between nearest neighbor
sites $i$ and $j$. For a uniform BEC with density $n_0=N_0/N_L$,
where $N_L$ is the number of lattice sites, one has
$\phi_{i0}^{*}\phi_{j0}=1/N_L$ such that the kinetic energy is given
by
\begin{equation}
E_{\mathrm{kin}}=-t\sum_{<ij>}\rho_{ij}=E^{\rm
BEC}_{\mathrm{kin}}+E^{\rm normal}_{\mathrm{kin}},
\end{equation}
where
\begin{equation}
E^{\rm BEC}_{\mathrm{kin}}=-t\sum_{<ij>}N_0, \;\;\;\; E^{\rm
normal}_{\mathrm{kin}}=-t\sum_{<ij>}\tilde{%
\rho}_{ij}.
\end{equation}
To derive a mean-field theory for lattice bosons via the limit of
high spatial dimensions the energy density $E_{\mathrm{kin}}/N_L$
needs to remain finite for $d$ or $Z\rightarrow \infty$. Since the
energy density of the condensate, $E^{\rm BEC}_{\mathrm{kin}}/N_L=Zt
n_0$, is proportional to $Z$ a non-trivial limit $Z\rightarrow
\infty$ is obtained only if the hopping amplitude is scaled as
$t=\tilde{t}/Z$, with $\tilde{t}$ = const.
\cite{freericks94,freericks96}  In the case of normal lattice bosons
(or fermions) the situation is characteristically different. Since
$\tilde{\rho}_{ij}$ is the transition amplitude for the hopping of a
boson from $j$ to one of the $Z$ neighboring sites $i$ the
respective normalized hopping \emph{probability} is
$|\tilde{\rho}_{ij}|^2 \propto 1/Z$, whence $\tilde{\rho}_{ij}\sim
1/\sqrt{Z}$. For the energy density of the normal bosons, $E^{\rm
normal}_{\mathrm{kin}}/N_L \propto Zt\tilde{\rho}_{ij}$, to remain
finite for $Z\rightarrow \infty$ the hopping amplitude must
therefore be scaled as in the case for fermions, namely, as
$t=\tilde{t}/ \sqrt{Z}$. \cite{metzner89} In the more general case
of hopping between sites $i$ and $j$ which are not nearest
neighbors, the amplitudes $t_{ij}$ have to be scaled as
\begin{equation}
t_{ij}=\tilde{t}_{ij}/(Z^{R_{ij}})^s,
\end{equation}
where $s=1$ (``integer scaling'') if the bosons are quantum
condensed and $s=1/2$ (``fractional scaling'') if they are in the
normal state.

The total energy of a single species ($\nu=1$) of correlated lattice
bosons described by the Hamiltonian (\ref{hamiltonian}) is given by
\begin{equation}
E=-t\sum_{< i,j>} N_0
-t\sum_{<i,j>}\tilde{\rho}_{ij}+\frac{1}{2}U\sum_i \langle
n_i(n_i-1)\rangle.  \label{full_energy}
\end{equation}

If the scaling of the hopping amplitudes in the limit
$Z\rightarrow\infty$ is performed on the level of the Hamiltonian
(or the energy $E$) two cases have to be distinguished:

(i) \underline{$N_0=0$}: In the absence of a BEC fractional
scaling ($t=\tilde{t}/\sqrt{Z}$) has to be employed to arrive at a
finite value of $E$ for $Z\rightarrow\infty$. We note that the
interaction is purely local and hence independent of the spatial
dimension of the system; consequently $U$ need not be scaled at all.

(ii) \underline{$N_0\ne 0$}: In this case integer scaling
($t=\tilde{t}/Z$) has to be employed. Thereby the contribution of
the condensate (first term in (\ref{full_energy})) remains. However,
at the same time the contribution of the non-condensed (normal)
bosons to the kinetic energy is suppressed $\propto 1/\sqrt{Z}$. The
normal bosons thus become immobile. It can be shown rigorously that
the mean-field equations obtained in this way are equivalent to
those derived by Fisher \textit{et al}. \cite{fisher89} which are
known to be exact in the limit of \emph{infinite-range hopping} (see
Section IV.C).

The above discussion shows that the construction of a mean-field
theory via the limit $Z\rightarrow\infty$ which is based on a
scaling of the hopping amplitudes in the Hamiltonian is either
restricted to the normal state, or removes the normal bosons from
the problem. This is unsatisfactory since the important
\emph{dynamical} coupling of normal and condensed bosons is then
eliminated from the outset. During the last 15 years the problem of
how to scale the hopping amplitudes without eliminating the normal
bosons presented an unsurmountable obstacle for the formulation of a
bosonic DMFT. Indeed, our discussion shows that such a theory cannot
be formulated on the level of a fully microscopic Hamilton operator,
i.e., without making an additional Bogoliubov mean-field type
assumption.\cite{Griffin98}

At this point it should be pointed out that there exists no \emph{a
priori} condition according to which the scaling of the hopping
amplitudes has to be performed in the Hamiltonian. Indeed, for the
free energy of the model to remain finite in the limit
$Z\rightarrow\infty$ the scaling can equally be performed in the
partition function (or in the action entering in the functional
integral which determines the partition function), from which the
free energy is calculated.

\section{Bosonic dynamical mean-field theory (B-DMFT)}

\subsection{General structure of the B-DMFT}

We will now show that the long-standing problem of the scaling of
the hopping amplitudes with the coordination number $Z$ or the
dimension $d$ can be resolved by considering the large dimensional
limit not in the Hamiltonian but in the \emph{action} determining
the Lagrangian density. Namely, integer scaling is employed whenever
a hopping amplitude is associated with anomalous expectation values
$\langle b_i \rangle$ and $\langle b_i^{\dagger} \rangle$, while
fractional scaling is used otherwise. The B-DMFT obtained in this
way treats normal and condensed bosons on equal footing and is thus
able to describe both phases, \emph{including the transition between
them}, in a thermodynamically consistent way.

\begin{figure}[bp]
\includegraphics [clip,width=13cm,angle=-00]{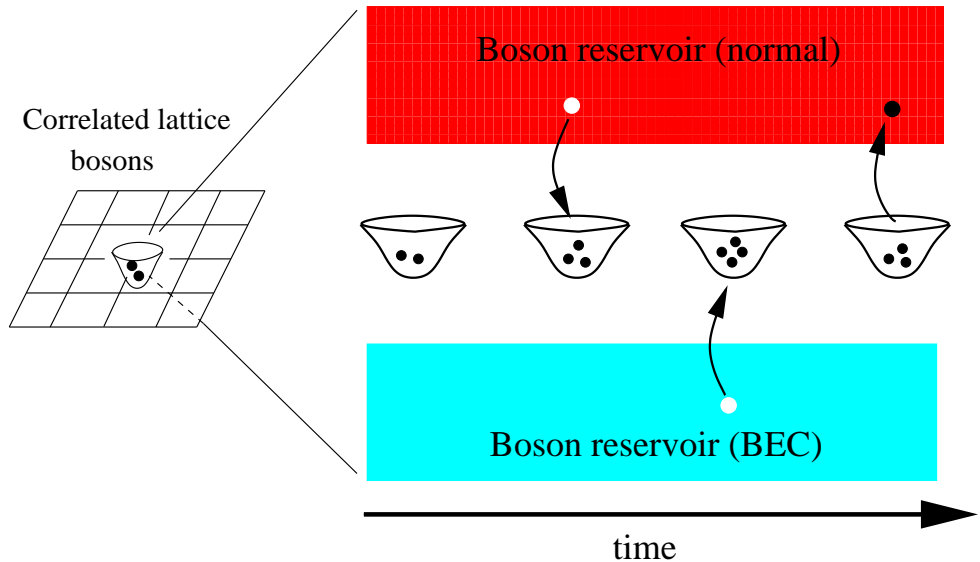}
\caption{Bosonic dynamical mean-field theory (B-DMFT): Within the B-DMFT the
full many-body lattice problem is reduced to a single-site problem which is
coupled to two reservoirs corresponding to bosons in the Bose-Einstein
condensate and in the normal state. The integer occupation of the site
changes in time and is determined by the local interactions and the
time-dependent properties of the particle reservoirs. Although the total
number of bosons is preserved, particles are scattered between the normal
and the condensate reservoirs via the single site as shown by arrows. This
schematic picture visualizes the idea of DMFT for lattice bosons in analogy
to the fermionic counterpart described in Ref.~\onlinecite{kotliar04}.}
\label{fig1}
\end{figure}

The general structure of the B-DMFT is shown in Fig.~(\ref{fig1}). In the
limit $d\rightarrow\infty$ the bosonic many-body lattice problem is mapped
onto a single-site problem with integer occupation. This site is coupled to
two particle reservoirs, one representing normal the other quantum condensed
bosons. These reservoirs represent the Weiss-type molecular fields of the
B-DMFT. Their properties are determined self-consistently by the B-DMFT
equations. Particles hop onto and off the site as a function of time, thus
changing the total number of bosons of the reservoirs. Therefore local
correlations lead to a dynamical depletion or filling of the condensate.

\subsection{B-DMFT equations}

Here we present the self-consistency equations of the B-DMFT for the
model (\ref{hamiltonian}); their derivation is discussed in detail
in Appendix A. The time evolution of $\nu$-bosons on the single site
$i=0$ is represented by the local propagator (Green function)
\begin{equation}
\mathbf{G}_{\nu}(\tau)=- \langle T_{\tau} \mathbf{b}_{\nu} (\tau)
\mathbf{b}_{\nu}^{\dagger}(0) \rangle_{S_{\mathrm{loc}}},
\label{dmft1}
\end{equation}
where we used the imaginary time, finite temperature formalism and Nambu
notation $\mathbf{b}_{\nu}^{\dagger}\equiv (b_{\nu}^*,b_{\nu})$; $T_{\tau}$
is the time ordering operator. The diagonal elements of the Green function
matrix in Nambu space represent the quantum-mechanical probability amplitude
for creating a boson on a site at one particular time and annihilating it
after a time $\tau$, or the time inverted process. The off-diagonal
elements, present only in the BEC phase, represent the amplitudes for
creating or annihilating two bosons at different times. In the path integral
formalism the probabilities of such processes are determined by the local
action, which in the B-DMFT is given by 
\begin{eqnarray}
S_{\mathrm{loc}}=-\int_0^{\beta}d\tau \int_0^{\beta}d\tau
^{\prime}\sum_{\nu} \mathbf{b}_{\nu}^{\dagger}(\tau)\;
\mathbf{\mathcal{G}}^{-1}_{\nu}(\tau-\tau^{\prime}) \;
\mathbf{b}_{\nu} (\tau ^{\prime}) + \int_0^{\beta} d\tau \sum_{\mu
\nu}\frac{U_{\mu\nu}}{2} n_{\mu}(\tau)[n_{\nu}(\tau)-\delta_{\mu\nu}
] + \int_0^{\beta}d \tau \sum_{\nu} \kappa_{\nu} \mathbf{\Phi
}^{\dagger}_{\nu}(\tau)\mathbf{b}_{\nu}(\tau) .  \label{dmft0}
\end{eqnarray}
Here $\kappa_{\nu}$ is a numerical factor depending on the lattice
structure, i.e., $\kappa_{\nu}\equiv \sum_{i\neq
0}\tilde{t}^{\nu}_{i0}/Z^{R_{i0}} $ for $d\rightarrow \infty$, and
$\kappa_{\nu}\equiv \sum_{i\neq 0}t^{\nu }_{i0}$ for an
approximation in finite dimensions.  The free ("Weiss"
\cite{georges96}) mean field propagator
$\mathbf{\mathcal{G}}^{-1}_{\nu}$, which is determined by the
properties of the reservoir of normal $\nu$-bosons, is related to
the local propagator $\mathbf{G}_{\nu}$ by the Dyson equation
\begin{eqnarray}
\mathbf{\mathcal{G}}^{-1}_{\nu}(i\omega_n)= \mathbf{G}_{\nu
}^{-1}(i\omega_n)+ \mathbf{\Sigma}_{\nu }(i\omega_n)\equiv
(i\omega_n\mathbf{\sigma}_3
-\mu_{\nu}\mathbf{1})-\mathbf{\Delta}_{\nu}(i\omega_n),
\label{Dyson_eq}
\end{eqnarray}
where $\mathbf{\mathcal{G}}$, and $\mathbf{\Sigma}$ are also
matrices in Nambu space.  Here $\omega_n=2\pi n/\beta$ are even
Matsubara frequencies with the inverse temperature $\beta=1/k_BT$
and $\mathbf{\Sigma}_{\mu}(i\omega_n) $ is the momentum-independent
(local) self-energy. The quantity $\mathbf{\Delta}_{\nu}$ describes
the resonant broadening of quantum-mechanical states on a lattice
site and may be interpreted as a hybridization of bosons on that
site with the surrounding bosonic bath. This hybridization function
is determined by the local correlations through eq.~(\ref{dmft0}).
The third term in (\ref{dmft0}) describes the coupling of a local
boson to the condensate, the latter being represented by an order
parameter $\mathbf{\Phi }^{\dagger}_{\nu}(\tau)$. In our formulation
this term arises naturally in the case of BEC and does not require a
Bogoliubov substitution. \cite{Griffin98}

The second B-DMFT equation is given by the lattice Hilbert transform
\begin{eqnarray}
\mathbf{G}_{\nu}(i\omega_n)= \sum_{\mathbf{k}} \left[(i\omega_n
\mathbf{\sigma}_3 -(\epsilon_{\mathbf{k}}^{\nu}
+\mu_{\nu})\mathbf{1} ) -\mathbf{\Sigma}_{\nu}(i\omega_n)
\right]^{-1},  \label{dmft2}
\end{eqnarray}
where $\epsilon_{\mathbf{k}}^{\nu}$ is the dispersion relation of
non-interacting $\nu$-bosons, $\mu_{\nu}$ the chemical potential,
$\mathbf{k} $ the wave vector, $\mathbf{1}$ a unity matrix, and
$\mathbf{\sigma}_3$ the Pauli matrix with $\pm 1$ on the diagonal.
\cite{georges96,pruschke95} Eqs. (\ref{dmft1},\ref{dmft2}) are the
counterparts to the self-consistency equations of the DMFT for
correlated lattice fermions. However, here these equations contain
the condensate wave function $\mathbf{\Phi}_{\nu}$, i.e., the order
parameter of the BEC, which enters as a source field in the action
(\ref{dmft1}). It can be determined exactly by calculating the
average
\begin{equation}
\mathbf{\Phi} _{\nu}(\tau) = \langle \mathbf{b}_{\nu}(\tau)
\rangle_{S_{\mathrm{loc}}}  \label{dmft3}
\end{equation}
together with eqs.~(\ref{dmft1}) and  (\ref{dmft2}). We note that in
equilibrium the time dependence enters via a trivial exponential
factor $e^{-\mu_{\nu}\tau}$ which can be eliminated by a gauge
transformation. \cite{agd}

Eqs.~(\ref{dmft1}-\ref{dmft3}) constitute the B-DMFT solution of the
generalized bosonic Hubbard model (\ref{hamiltonian}). These
equations are exact in the $d\rightarrow \infty$ limit and provide a
comprehensive, thermodynamically consistent and conserving
approximation in finite dimensions. In other words, the B-DMFT
derived here is the first mean-field theory for correlated lattice
bosons which has all the attractive features characterizing the now
well-established fermionic DMFT. \cite{georges96,kotliar04} In
particular, the B-DMFT can be expected to be the best approximation
to many-boson problems with strong local correlations since the
on-site quantum fluctuations of the spin or density are treated
exactly. Spatial correlations are neglected but can be restored,
e.g., within cluster extensions of the B-DMFT. Furthermore,
long-range ordered phases can be described within the B-DMFT by
properly choosing the self-consistency conditions in analogy with
the fermionic case. \cite{georges96}

\subsection{B-DMFT and Gross-Pitaevskii equation}

The exact Euler-Lagrange equation of motion for the field $b(\tau)$
is obtained from the stationary conditions $\delta
S_{\mathrm{loc}}[{\ b}^{\dagger}_{\nu}, {\ b}_{\nu}]/\delta {\
b}_{\nu}^{\dagger}=0$ of the local B-DMFT action and is given by
\begin{eqnarray}
\partial_{\tau}{\ b}_{\nu}(\tau)-\int_0^{\beta} d\tau^{\prime}
[ {\ \Delta}_{\nu }^{11}(\tau-\tau^{\prime}) {\ b}_{\nu}
(\tau^{\prime})+ {\ \Delta}_{\nu }^{12}(\tau-\tau^{\prime}) {\
b}_{\nu}^{\dagger} (\tau^{\prime}) ]+ \kappa_{\nu} {\
b}_{\nu}(\tau)+ \sum_{\mu} U_{\mu \nu}{\ b}_{\mu}^{\dagger}(\tau){\
b}_{\mu}(\tau) {\ b}_{\nu}(\tau) =\mu_{\nu}{\ b}_{\nu}(\tau).
\label{dmft33}
\end{eqnarray}
If one replaces each field $b(\tau)$ by its expectation value (the order
parameter ${\ \Phi}_{\nu}$) one arrives at 
\begin{eqnarray}
\partial_{\tau}{\ \Phi}_{\nu}(\tau)-\int_0^{\beta} d\tau^{\prime}[ {\ \
Delta}_{\nu }^{11}(\tau-\tau^{\prime}) {\ \Phi}_{\nu}
(\tau^{\prime}) + {\ \Delta}_{\nu }^{12}(\tau-\tau^{\prime}) {\
\Phi}_{\nu}^* (\tau^{\prime})] + \kappa_{\nu} {\Phi}_{\nu}(\tau)+
\sum_{\mu} U_{\mu \nu}|{\ \Phi}_{\mu}(\tau)|^2{\ \Phi}_{\nu}(\tau)
=\mu_{\nu}{\ \Phi}_{\nu}(\tau). \label{dmft333}
\end{eqnarray}
This is a generalization of the standard Gross-Pitaevskii mean-field
equation, a non-linear differential equation for the condensed
bosons which can be derived within the time-dependent Hartree-Fock
approximation. \cite{Griffin98} At present it is not clear whether
the replacement $b_{\nu}(\tau) \rightarrow \Phi_{\nu}(\tau)$, i.e.,
the factorization of the correlation function $\langle {\
b}_{\mu}^{\dagger}(\tau){\ b}_{\mu}(\tau) {\ b}_{\nu}(\tau)\rangle =
|{\ \Phi}_{\mu}(\tau)|^2{\ \Phi}_{\nu}(\tau)$, holds rigorously as
in other mean-field theories. \cite{gp,gp1,gp2,gp3}
Eq.~(\ref{dmft333}) is the \emph{classical} equation of motion of
the condensate of lattice bosons in the $d\rightarrow \infty$ limit.
The second term on the l.h.s. of (\ref{dmft333}) describes the
retarding effects of normal bosons on the condensate and makes the
generalized Gross-Pitaevskii equation a non-linear
integro-differential equation. We note that in the standard
Gross-Pitaevskii equation the hybridization term
$\mathbf{\Delta}_{\nu}$ is missing. Once $\mathbf{\Delta}_{\nu} $
has been determined by solving the B-DMFT self-consistency equations
(\ref{dmft333}) can be used to determine $\mathbf{\Phi}_{\nu}(\tau)$
and then calculate any response function of the condensate due to
external perturbations, e.g., to describe Bragg scattering. In
equilibrium ${\mathbf{\Phi}}_{\nu}$ can be expected to be
independent of $\tau$. Then the stationary solution of
(\ref{dmft333}) is easily obtained by solving a set of linear
equations. For example, for the spinless bosonic Hubbard model (no
index $\nu$) we obtain $|\Phi|^2=[\mu-\kappa +
\Delta^{11}(\omega_n=0)+\Delta^{12}(\omega_n=0)]/U$. The condensate
density depends explicitly on the zero mode components of the
hybridization functions for the normal subsystem.

\section{The B-DMFT in different limits of parameter space}

\begin{figure}[tbp]
\includegraphics [clip,width=13cm,angle=-00]{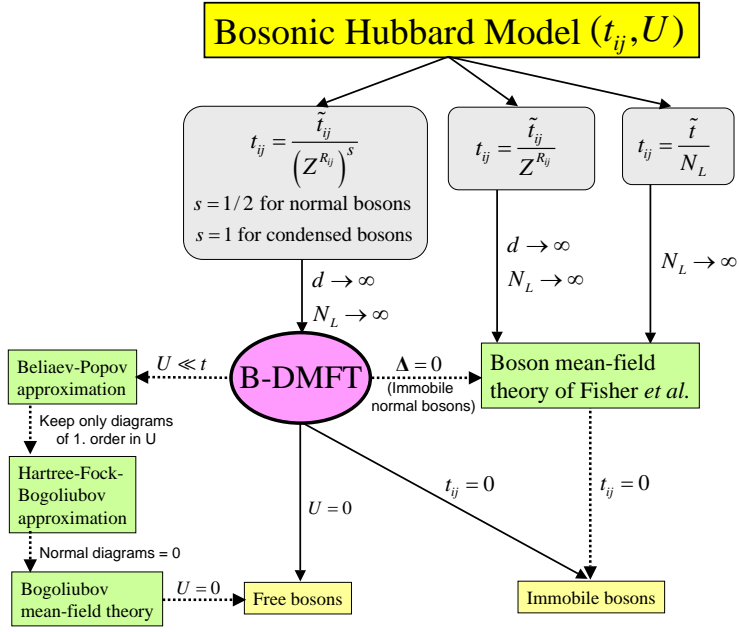}
\caption{Relation of the B-DMFT to other approximations and exact limits;
see Sec. IV. }
\label{fig4}
\end{figure}

The B-DMFT is a comprehensive mean-field theory for correlated
bosons on a lattice, meaning that the theory is valid for all input
parameters and temperatures. This is an essential prerequisite for
obtaining a reliable, approximate description of those parts of the
phase diagram which cannot be studied perturbatively. Above all, the
B-DMFT reproduces all known results obtained in special limits of
the parameter space as depicted in Fig.~{\ref{fig4}}. In particular,
the exactly solvable limits of free and immobile bosons,
respectively, and of well-known static mean-field approximations can
be obtained directly from the B-DMFT. In the following we discuss
these limits in details.

\subsection{Free bosons}

In the non-interacting limit, $U_{\mu \nu}=0$, the problem is trivially
solvable in all dimensions. Since there is no interaction, all correlation
functions factorize and the cavity method employed in Appendix A becomes
exact. In this case the BEC is described within the grand-canonical ensemble
by a non-vanishing order parameter $\Phi$. However, the off-diagonal Green
(hybridization) functions of the normal bosons are zero. Explicitly, the
local action has a bilinear (Gaussian) form
\begin{eqnarray}
S_{\mathrm{loc}}^{\mathrm{non-interacting}}= -\int_0^{\beta}d\tau
\int_0^{\beta}d\tau ^{\prime}\sum_{\nu} \mathbf{b}_{\nu}^{\dagger}(\tau)\;
\mathbf{\mathcal{G}}^{-1}_{\nu}(\tau-\tau^{\prime}) \; \mathbf{b}_{\nu}
(\tau ^{\prime}) + \int_0^{\beta}d \tau \sum_{\nu} \kappa_{\nu} \mathbf{\Phi
}^{\dagger}_{\nu}(\tau)\mathbf{b}_{\nu}(\tau),
\end{eqnarray}
with $\mathcal{G}^{-1}_{\nu} (\tau-\tau^{\prime}) = (-\partial_{\tau}
\mathbf{\sigma}_3 + \mu_{\nu}\mathbf{1}) \delta(\tau-\tau^{\prime}) -
\mathbf{\ \Delta}_{\nu}(\tau-\tau^{\prime})$. All known equations, e.g.,
that for the particle number or the compressibility, can be easily derived
from this action (see Appendix B). We note that in the non-interacting limit
the hopping amplitudes need not be scaled for $d<\infty$. However, to obtain
a meaningful limit $d\rightarrow \infty$ the scaling scheme introduced in
this paper is necessary. Otherwise the condensate and normal bosons would
not be treated on equal footing. That is, if only fractional scaling is
employed, one obtains spurious infinities in the condensate phase, whereas
if only integer scaling is used, the normal bosons (which contribute
significantly at temperatures close or above $T_{\mathrm{BEC}}$) become
immobile.

\subsection{Immobile bosons (``atomic limit'')}

In the atomic limit, $t_{ij}^{\nu}=0$, all lattice sites are
decoupled and the particles are immobile. In this case the order
parameter $\mathbf{\Phi }_{\nu}=0$ since no condensation is
possible, and also the hybridization function
$\mathbf{\Delta}_{\nu}=0$. For arbitrary dimensions the exact action
is then given by a sum over all equivalent sites with the same local
action
\begin{eqnarray}
S_{\mathrm{loc}}^{\mathrm{atomic}}=-\int_0^{\beta}d\tau \sum_{\nu}
\mathbf{b}_{\nu}^{\dagger}(\tau)\; (-\partial_{\tau}
\mathbf{\sigma}_3 + \mu_{\nu}\mathbf{1}) \; \mathbf{b}_{\nu} (\tau )
\int_0^{\beta} d\tau \sum_{\mu \nu}\frac{U_{\mu\nu}}{2}
n_{\mu}(\tau)[n_{\nu}(\tau)-\delta_{\mu\nu} ] , \label{atom}
\end{eqnarray}
as obtained from the action (\ref{dmft0}) with $\mathbf{\ \Phi
}_{\nu}=0$ and $\mathbf{\Delta}_{\nu}(\tau)=0$ in
$\mathcal{G}^{-1}_{\nu} (\tau-\tau^{\prime}) = (-\partial_{\tau}
\mathbf{\sigma}_3 + \mu_{\nu}\mathbf{1}) \delta(\tau-\tau^{\prime})
- \mathbf{\Delta}_{\nu}(\tau-\tau^{\prime})$.

\subsection{Mean-field theory of Fisher \textit{et al}. [\onlinecite{fisher89}]}

The mean-field field theory of Fisher \textit{et  al}.
\cite{fisher89} is known to be the exact solution of the bosonic
Hubbard model when the hopping amplitude is independent of distance
and is scaled with the number of lattice sites $N_L$, i.e.
$t_{ij}=\tilde{t}/N_L$. \cite{fisher89,bru03} This is also called
the limit of ``infinite-range hopping''.
\cite{fisher89,bru03,dongen89} The free energy density of the
bosonic Hubbard model in this limit has the form
\begin{eqnarray}
F^{\mathrm{infinite-range}}=F_{\mathrm{at}} -kT\ln \langle T_{\tau}
e^{-\int_0^{\beta}d\tau \sum_{\nu} \kappa_{\nu} \mathbf{\Phi
}^{\dagger}_{\nu}\mathbf{b}_{\nu}(\tau)} \rangle_{H_{\mathrm{at}}}
-\sum_{\nu}\kappa_{\nu}|\Phi_{\nu}|^2.  \label{smft3}
\end{eqnarray}
Here the atomic part $F_{\mathrm{at}} = -kT \ln \mathrm{Tr} \exp
(-\beta H_{\mathrm{at}})$ is given by the Hamiltonian
$H_{\mathrm{at}}$, which is obtained from the lattice Hamiltonian by
setting all hopping amplitudes equal to zero. The average in the
second term is taken with respect to $H_{\mathrm{at}}$. The
stationarity condition for the free energy (\ref{smft3}) with
respect to $\Phi_{\nu}$, i.e. $\partial
F^{\mathrm{infinite-range}}/\partial \mathbf{\Phi}_{\nu} =0$ then
yields the self-consistent mean-field equation
\begin{eqnarray}
\kappa_{\nu}\mathbf{\Phi}_{\nu}=kT\frac{\langle T_{\tau}
\mathbf{b}_{\nu} e^{-\int_0^{\beta}d\tau \sum_{\nu} \kappa_{\nu}
\mathbf{\Phi }^{\dagger}_{\nu}\mathbf{b}_{\nu}(\tau)}
\rangle_{H_{\mathrm{at}}}}{\langle T_{\tau} e^{-\int_0^{\beta}d\tau
\sum_{\nu} \kappa_{\nu} \mathbf{\Phi
}^{\dagger}_{\nu}\mathbf{b}_{\nu}(\tau)} \rangle_{H_{\mathrm{at}}}}.
\label{Fisher_mf_eq}
\end{eqnarray}
Since the normal bosons are immobile, and thus not dynamically coupled to
the condensed bosons, the theory of Fisher \textit{et al}. \cite{fisher89}
is a \emph{static} mean-field theory.

We now show that the mean-field equation (\ref{Fisher_mf_eq}) can
also be obtained in the $d\rightarrow \infty $ limit as put forward
in Refs.~\onlinecite{rok,freericks94,freericks96}. Indeed, by
employing the cavity method  for lattice bosons and using only
integer scaling for the hopping amplitudes the local action in the
$d\rightarrow \infty$ ($Z\rightarrow \infty$) limit takes the form
\begin{eqnarray}
S_{\mathrm{loc}}^{d\rightarrow\infty, \;\mathrm{integer\;scaling}}=
-\int_0^{\beta}d\tau \sum_{\nu} \mathbf{b}_{\nu}^{\dagger}(\tau)\;
(-\partial_{\tau} \mathbf{\sigma}_3 + \mu_{\nu}\mathbf{1}) \;
\mathbf{b}_{\nu} (\tau )  \nonumber \\
+ \int_0^{\beta} d\tau \sum_{\mu \nu}\frac{U_{\mu\nu}}{2}
n_{\mu}(\tau)[n_{\nu}(\tau)-\delta_{\mu\nu} ] + \int_0^{\beta}d \tau
\sum_{\nu} \kappa_{\nu} \mathbf{\Phi
}^{\dagger}_{\nu}(\tau)\mathbf{b}_{\nu}(\tau) .  \label{smft1}
\end{eqnarray}
This expression differs from the local action in the atomic limit,
(\ref{atom}), by the presence of the last term which describes the
condensate.However, in equilibrium the BEC order parameter is time
independent in which case (\ref{smft1}) yields the free energy
density as
\begin{eqnarray}
F^{d\rightarrow\infty, \;\mathrm{integer\;scaling}}=F_{\mathrm{at}}
-kT\ln \langle T_{\tau} \exp\left(-\int_0^{\beta}d\tau \sum_{\nu}
\kappa_{\nu} \mathbf{\Phi
}^{\dagger}_{\nu}\mathbf{b}_{\nu}(\tau)\right)
\rangle_{H_{\mathrm{at}}}.  \label{smft2}
\end{eqnarray}
The BEC order parameter $\mathbf{\Phi}_{\nu}$ obeys the self-consistent
equation
\begin{eqnarray}
\mathbf{\Phi}_{\nu} = \frac{1}{Z_{\mathrm{loc}}^{d\rightarrow\infty,
\;\mathrm{integer\;scaling}}} \cdot \int D [b,b^*] \;
\mathbf{b}_{\nu}\; e^{-S_{\mathrm{loc}}^{d\rightarrow\infty,
\;\mathrm{integer\;scaling}}[b,b^*;\mathbf{\Phi}_{\nu} ]}.
\label{BEC_OP}
\end{eqnarray}
Other correlation functions and observables can be determined similarly.

The free energy density (\ref{smft2}) is seen to differ from
(\ref{smft3}) only by the absence of the last term proportional to
the density of the condensate. Nevertheless, the \emph{equations}
for the BEC order parameter as well as correlation functions and
observables are the same. Thus the approximation of constant hopping
amplitude (``infinite-range hopping'') and the $d\rightarrow \infty
$ limit with integer scaling give rise to the same mean-field
equations. At $T=0$ these equations can also be derived by yet
another approximation, namely a variational method using a
Gutzwiller-type wave function. \cite{rok,krauth} The mean-field
theory of Fisher \textit{et al}.\cite{fisher89} and its
generalization to spinful bosons were widely used to investigate
quantum phase transitions and the phase diagrams of correlated
lattice boson systems and of mixtures of lattice bosons and
fermions.\cite{lewenstein06,Bloch07,jaksch98,sachdev02,Isacsson05,Huber07,Sachdev,Altman,Tit}

Eq. (\ref{smft2}) and (\ref{BEC_OP}) can also be obtained directly
from the B-DMFT self-consistency equations by neglecting the
hybridization function, i.e. by setting $\mathbf{\Delta}_{\nu}=0$.
Then the local action of the B-DMFT, (\ref{dmft0}), is the same as
that in (\ref{smft1}). Furthermore, (\ref{dmft2}) is satisfied
automatically in this limit since only the state with $\mathbf{k}=0$
is taken into account. It should be noted, however, that in the
absence of the hybridization function $\mathbf{\Delta}_{\nu}$ the
non-interacting limit of the normal bosons cannot be reproduced,
i.e., the mean-field theory of Fisher \textit{et  al}.
\cite{fisher89} does not describe the limit of free, normal  bosons.

\subsection{Weak-coupling (Bogoliubov) mean-field theory}

A perturbation expansion to first order in $U_{\mu\nu}$ is
equivalent to the Hartree-Fock-Bogoliubov approximation with the
static self-energy $\Sigma_{\nu}^{11}=2\sum_{\mu} U_{\nu\nu}
\bar{n}^{\mathrm{BEC}}_{\mu}-2 \sum_{\omega_n,\mu}
U_{\nu\mu}\mathcal{G}_{\mu}(\omega_n)/\beta$ and $\Sigma_{\nu}^{12}
=\sum_{\mu}U_{\nu\mu} \bar{n}^{\mathrm{BEC}}_{\mu}$. For such a
self-energy the self-consistency condition (\ref{dmft2}) is
equivalent to the self-consistent Hartree-Fock-Bogoliubov
approximation (sometimes called ``first-order Popov''
approximation). \cite{Griffin98,agd} This self-consistent
approximation is known to lead to a gapped spectrum in the condensed
phase because off-diagonal elements in the self-energy are
calculated in higher order due to self-consistency. By contrast, the
Bogoliubov approximation \cite{Griffin98,agd} is obtained if only
particular diagrams corresponding to the self-energies
$\Sigma_{\nu}^{11}=2\sum_{\mu}U_{\nu\mu}
\bar{n}^{\mathrm{BEC}}_{\mu}$ and $\Sigma_{\nu}^{12}
=\sum_{\mu}U_{\nu\mu} \bar{n}^{\mathrm{BEC}}_{\mu}$ are taken into
account.

The second-order expansion contains many diagrams, see
Refs.~\onlinecite{Griffin98,agd}. Checking term by term we find that
the B-DMFT reproduces the Beliaev-Popov approximation (sometimes
called ``second-order Popov approximation'') \cite{Griffin98} if, in
addition, in the latter approximation only local irreducible
self-energy diagrams (consistent with the $d\rightarrow \infty$
limit) are retained.

\section{B-DMFT Solution of the Bosonic Falicov-Kimball model}

We now apply the B-DMFT to study BEC in a mixture of two different
species of bosons: itinerant $b$-bosons and immobile $f$-bosons. We
assume the $b$-bosons not to interact with each other but only with
$f$-bosons, while f-bosons interact also mutually, i.e., $U_{bb}=0$,
$U_{bf}>0$, $U_{ff}>0$ in (\ref{hamiltonian}). We call this the
bosonic Falicov-Kimball (BFK) model since it is a bosonic
generalization of the Falicov-Kimball model for fermions which has
been widely studied in condensed-matter physics. \cite{freericks03}
Experimentally such a system can be realized by loading an optical
lattice either with a mixture of two different species of bosonic
alkali atoms (e.g., $^7$Li and $^{87}$Rb), or by one kind of atom
with two different hyper-spin states (i.e., with $F=1,2$ and
specific values of the $z $-component of $F$). In addition, the
electric fields generating the potentials of the optical lattice and
the external magnetic field controlling the Feshbach resonances
should be tuned such that one species of particles is immobile and
the other is non- (or only weakly) interacting. The realization of a
fermionic Falicov-Kimball model by cold fermionic atoms in an
optical lattice was discussed in
Refs.~\onlinecite{Ziegler,Freericks2008}.

It is important to note that, in spite of the immobility of the $f$-bosons,
the BFK model is still a many-body problem because the immobile particles
are thermodynamically coupled to the mobile particles by the interaction. In
particular, the optimal configuration of the localized bosons depends on the
interaction, temperature and density of the particles. In the fermionic
counterpart one finds that the position of the immobile particles is either
random or long-range ordered; phase separation between these two components
can also occur. The numerical solution of the Falicov-Kimball model is
limited to small lattices and requires an annealed average over a large
number of configurations of immobile particles.

For the BFK model the local impurity problem can be integrated
analytically. The self-consistency equations can be then solved  by
standard numerical techniques. Since the $f$-bosons are immobile,
their number on each site is conserved. Hence, the $f$-boson
subsystem cannot undergo BEC and the occupation number operator
$n_f$ of the single site becomes a classical variable with
$n_f=0,1,2,..$. The local action (\ref{dmft1}) is then quadratic in
the bosonic operators. Consequently, the local propagator
$G_{b}(i\omega_n)$ for $b$-bosons and the local partition function
$Z_{\mathrm{loc}}(\mu_b,\mu_f)$, and thereby the BEC transition
temperature $T_{\mathrm{BEC}}$ for the $b$-bosons, can be evaluated
directly. The local partition function of the BFK model is
determined by
\begin{equation}
Z_{\mathrm{loc}}(\mu_b,\mu_f)=\sum_{n_f=0,1,2,..}^{\infty} e^{\beta n_f(
\mu_f -\frac{U_{ff}n_f}{2})}Z_{\mathrm{loc}}^0(\mu_b-U_{bf}n_f,\mu_f),
\end{equation}
where
\begin{equation}
Z^0_{\mathrm{loc}}(\mu_b,\mu_f)\sim
e^{-\frac{\kappa_b|\phi_b|^2}{\mu_b-\Delta_{b}(0)}}\prod_{\omega_n}
\left( \frac{1}{i\omega_n+\mu_b-\Delta_{b}(i\omega_n)}\right)
\end{equation}
is the partition function for $U_{bf}=0$. The local propagator for
normal $b$-bosons is given by
\begin{eqnarray}
G_{b}(i\omega_n)=\sum_{n_f=0,1,2,...}^{\infty}
\frac{w_{n_f}}{i\omega_n+\mu_b-U_{bf}n_f-\Delta_{b}(i\omega_n)}.
\label{green_fk}
\end{eqnarray}
Here
\begin{equation}
w_{n_f}=e^{\beta n_f (\mu_f
-\frac{U_{ff}n_f}{2})}\frac{Z_{\mathrm{loc}}^0(\mu_b-U_{bf}n_f,\mu_f)}{Z_{\mathrm{loc}}(\mu_b,
\mu_f) } \label{probability}
\end{equation}
is the probability for the single site to be occupied by exactly
$n_f=0,1,2,...$ bosons. For hard-core $f$-bosons ($U_{ff}=\infty$)
this leads to $w_{n_f=1}=\bar{n}_f$ and $w_{n_f=0}=1-\bar{n}_f$. In
this case $\bar{n}_f $ rather than $\mu_f$ is used as an independent
thermodynamical variable. The propagator (\ref{green_fk}) describes
quantum and thermal fluctuations of normal bosons. In the absence of
the interaction between $b$-bosons, off-diagonal terms in the local
propagator are zero. The Gross-Pitaevskii equation is then obviously
exact and reduces to a homogeneous, linear equation of the form
$[-i\omega_n-\mu_b+\kappa_b+\Delta_{b}(i\omega_n)]\phi_b(i\omega_n)=0$
for each Fourier component. For $\omega_n\neq 0$ the only solution
is $\phi_b(\omega_n\neq 0)=0$. The static ($\omega_n=0$) component
of the BEC order parameter is finite if
$\mu_b=\kappa_b+\Delta_{b}(0)$ and must be determined by fixing the
average density of $b$-bosons.

\begin{figure}[tbp]
\includegraphics
[clip,width=13cm,angle=-00]{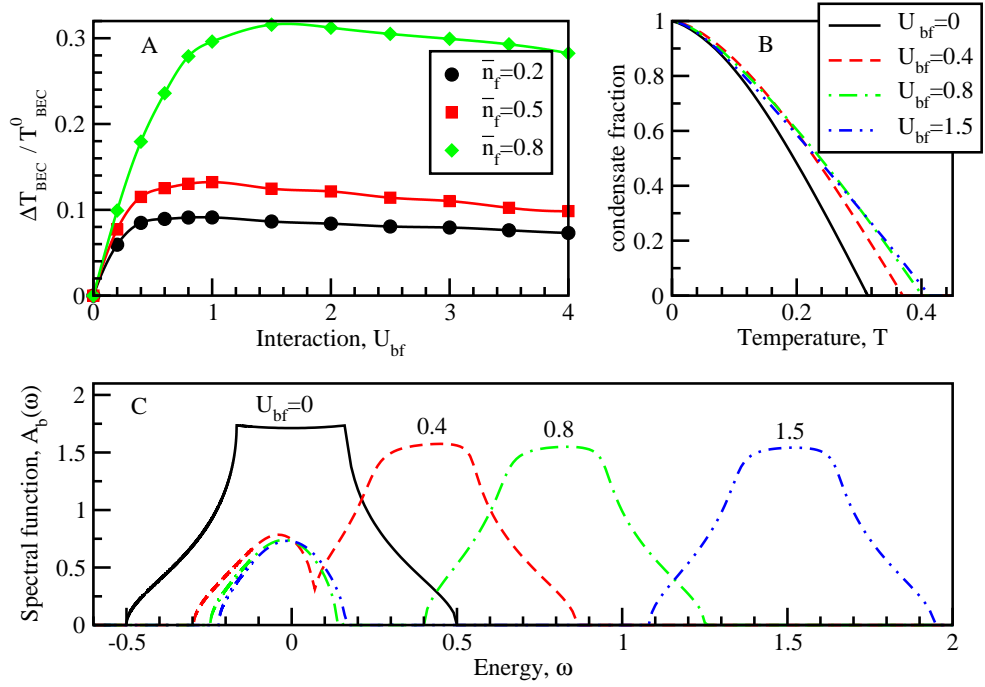} \caption{Bose-Einstein
condensation of a mixture of itinerant and localized, correlated
lattice bosons: (A) Enhancement of the BEC transition temperature
with increasing interaction strength $U_{bf}$ in a two component
boson mixture with different densities $\bar{n}_f$ of the localized
$f$-bosons. (B) Dependence of the condensate fraction
$\bar{n}_b^{\mathrm{BEC}}(T)/\bar{n}_b$ on temperature for different
interactions $U_{bf}$ at $\bar{n}_f=0.8$. (C) Spectral functions for
different values of $U_{bf}$ at $\bar{n}_f=0.8$. The increase of
$T_{BEC}$ and the condensate fraction with increasing $U_{bf} $ and
$\bar{n}_f$ is caused by correlation effects leading to a
redistribution of the spectral weight for the $b$-bosonic subsystem.
The correlation gap opens when $U_{bf}$ exceeds a critical value
which depends on $\bar{n}_f$. The opening of the gap is not
associated with a phase transition of the mobile bosons. Results are
obtained for a three-dimensional cubic lattice with unit band-width
and $\bar{n}_b=0.65$. In the hard-core limit the spectral functions
are temperature independent because the occupation probability of
$f$-bosons is either $\bar{n}_f$ or $1-\bar{n}_f$.} \label{fig2}
\end{figure}

\begin{figure}[tbp]
\includegraphics [clip,width=13cm,angle=-00]{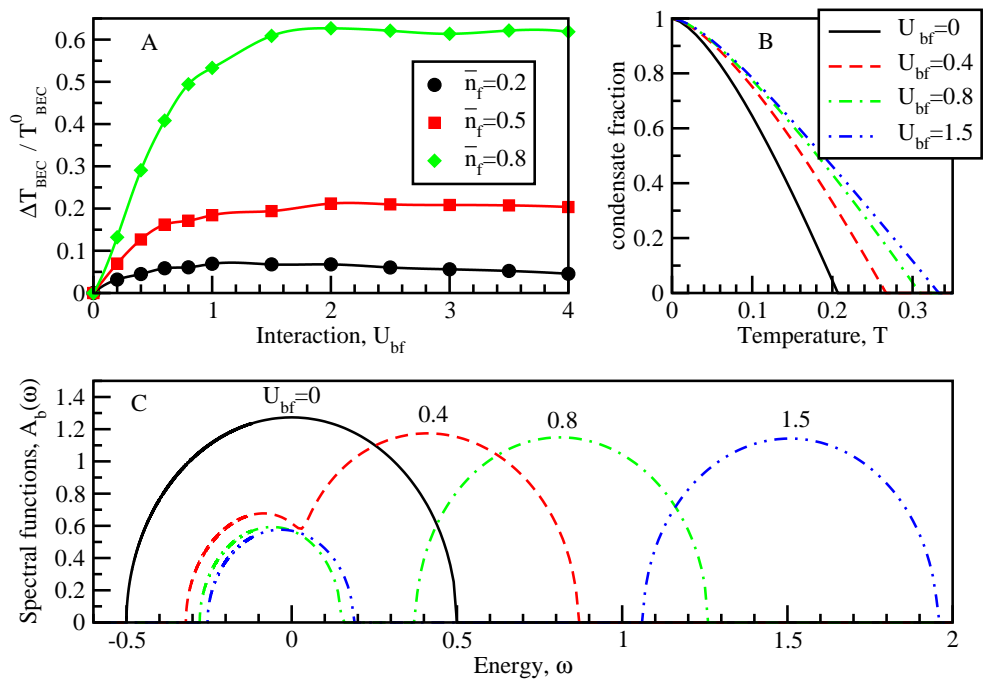}
\caption{Bose-Einstein condensation of a mixture of itinerant and
localized, correlated lattice bosons: In contrast to Fig.~2 results
here are obtained for a Bethe lattice with infinite $Z$, unit
band-width, and $\bar{n}_b=0.5$. They are exact for the
bosonic-Falicov-Kimball model. (A) Enhancement of the BEC transition
temperature with increasing interaction strength $U_{bf}$ in a two
component boson mixture with different densities $\bar{n}_f$ of the
localized $f$-bosons. (B) Dependence of the condensate fraction
$\bar{n}_b^{\mathrm{BEC}}(T)/\bar{n}_b$ on temperature for different
interactions $U_{bf} $ at $\bar{n}_f=0.8$. (C) Spectral functions
for different values of $U_{bf}$ at $\bar{n}_f=0.8$. The increase of
$T_{BEC}$ and the condensate fraction with increasing $U_{bf}$ and
$\bar{n}_f$ is caused by correlation effects leading to a
redistribution of the spectral weight for the $b$-bosonic subsystem.
} \label{fig3}
\end{figure}

A striking result obtained for this model is an \emph{enhancement}
of $T_{\mathrm{BEC}}$ for increasing repulsion between the $b$- and
$f$-bosons, with a maximum of $T_{\mathrm{BEC}}$ at intermediate
values of $U_{bf}$ and a saturation at large $U_{bf}$. This behavior
is explicitly seen in Fig.~\ref{fig2}A for hard-core $f$-bosons
($U_{ff}=\infty$, i.e., $n_f=0,1$) on a simple cubic lattice where
we plotted the relative change of $T_{\mathrm{BEC}}$ with respect to
$T_{\mathrm{BEC}}^0$ in the non-interacting system. The increase of
$T_{\mathrm{BEC}}$ is due to the blocking of a fraction of sites by
heavy atoms which increases the density of the $b$-bosons. However,
this argument cannot explain the non-monotonicity of $T_{BEC}$ vs
$U_{bf}$ shown in Fig.~\ref{fig2}A. In fact, the maximum is due to
the correlation induced band splitting and the  narrowing of the
lower subband (see Fig.~\ref{fig2}C) which lead to  an increase and
decrease of $T_{\mathrm{BEC}}$, respectively.  Furthermore, at fixed
temperature $T$ the average condensate density
$\bar{n}_b^{\mathrm{BEC}}(T)$ is also found to increase with
increasing repulsion (see Fig.~\ref{fig2}B), although the
interaction induced scattering between bosons usually removes
particles from the condensate, thereby \emph{reducing} its density.
\cite{martin05} At zero temperature all $b$-bosons are in the
condensate. Similar results are found for other lattices, Cf.
Fig.~\ref{fig3}.

These results originate from local correlations which are captured exactly
by the B-DMFT, but not by conventional approximations. The consequences can
be inferred by considering the total density of $b$-bosons
\begin{equation}
\bar{n}_b=\bar{n}_b^{\mathrm{BEC}}(T)+\int d\omega\;
\frac{A_{b}(\omega+\mu_b)}{\exp(\omega/T)-1}.
\end{equation}
The second term gives the contribution of normal $b$-bosons for
which the spectral function
$A_{b}(\omega)=-\mathrm{Im}G_{b}(\omega)/\pi$ is shown in
Fig.~\ref{fig2}C for different $U_{bf}$ values. The spectral weight
is seen to be strongly redistributed, forming lower- and
upper-Hubbard subbands at low and high energies $\omega$,
respectively, which are separated by the energy $U_{bf}$. We note
that the splitting and rounding of the shapes are genuine
correlation effects. The occupation probability of normal
$b$-bosons, i.e., the Bose-Einstein distribution function, decays
exponentially with increasing $\omega$. This implies an extremely
small particle number in the upper Hubbard subband at large
$U_{bf}$. Since the total number of $b$-bosons is constant, the
particles are necessarily transferred into the condensate. Hence, at
fixed temperature $\bar{n}_b^{\mathrm{BEC}}$ \emph{increases},
thereby enhancing $T_{\mathrm{BEC}}$. The spectral weight
contributing to the upper Hubbard subband is also proportional to
$\bar{n}_f$, implying that $T_{\mathrm{BEC}}$ increases with
$\bar{n}_f$, too.

The B-DMFT prediction of an increasing $T_{\mathrm{BEC}}$ and
condensate density due to local correlations are expected to be
observable in mixtures of mobile and localized bosons on three
dimensional optical lattices. Such correlations can thus be employed
in the laboratory to enhance $T_{\mathrm{BEC}}$ of bosonic
condensates. We also note that on bipartite lattices with special
densities of bosons, e.g. $\bar{n}_f=\bar{n}_b=0.5$, long-range
order in the density of the $f$-subsystem and, in turn, a supersolid
phase in the $b$-subsystem is expected to form. Obviously the
physics of this seemingly simple bosonic model is extraordinarily
rich.

\section{Conclusions}

In this paper we derived the first comprehensive, thermodynamically
consistent theoretical framework for the investigation of correlated lattice
bosons --- a bosonic dynamical mean-field theory (the B-DMFT). In analogy to
its fermionic counterpart the B-DMFT becomes exact in the limit of high
spatial dimensions $d$ or coordination number $Z$ and may be employed to
compute the phase diagram and thermodynamics of interacting lattice boson
systems in the entire range of microscopic parameters. The B-DMFT requires a
different scaling of the hopping amplitude with $Z$ depending on whether the
system is in the normal or the Bose-condensed phase. This additional
difficulty compared to the fermionic case prevented the formulation of the
B-DMFT in the past. As shown here it can be overcome by performing the
scaling not in the Hamiltonian but in the action. The B-DMFT equations
consist not only of a bosonic single-impurity problem in the presence of a
self-consistency condition (the momentum integrated Dyson equation), but
involve an additional coupling to the condensate wave function.

We documented the comprehensive nature of the B-DMFT by explicitly
reproducing results previously obtained in special limits of
parameter space and by deriving other bosonic mean-field theories.
For example, by calculating the local-self energy of the B-DMFT in
the weak coupling regime, $U/t \ll 1$, in perturbation theory and by
including all normal and anomalous terms to first order in $U/t$ one
obtains the Hartree-Fock-Bogoliubov self-consistent mean-field
approximation.\cite{Griffin98} By neglecting the anomalous terms the
standard Bogoliubov theory for lattice bosons is recovered.
Inclusion of the second-order corrections to the local self-energy
corresponds to the Beliaev-Popov approximation (with the additional
assumption of a local self-energy). Furthermore, by neglecting all
terms containing the hybridization function in the local action one
obtains the mean-field theory developed in
Ref.~\onlinecite{fisher89} and
Refs.~\onlinecite{freericks94,freericks96}, which corresponds to the
exact solution of the bosonic Hubbard model (\ref{hamiltonian})  in
the large dimension limit if only integer scaling is applied.

In contrast to previous mean-field theories the B-DMFT constructed
here treats normal and condensed bosons on equal footing. In
particular, the B-DMFT takes into  account effects due to finite
hopping and dynamical broadening of the quantum levels. The
inclusion of the hybridization function leads to important changes
of the results of the static  mean-field theory. For example, the
Hubbard $\delta$-peaks in the  spectral functions now acquire a
finite width, and the Mott transition occurs already when these
bands start to overlap. We note that the B-DMFT is not merely a
perturbative improvement of the static mean-field theory
\cite{fisher89,freericks94,freericks96,kampf93} with respect to the
hybridization since in the B-DMFT the hybridization function is
included to all orders.

We applied the B-DMFT to solve the bosonic Falicov-Kimball model, i.e., a
lattice model of itinerant and localized, interacting bosons. Due to the
localized nature of the interacting bosons (providing a type of annealed
disorder to the system since the localized particles are thermodynamically
coupled to the itinerant bosons) the problem reduces to a set of algebraic
equations. We find that local correlations \emph{enhance} the transition
temperature into the condensate and can thus be employed in the laboratory
to increase $T_{\mathrm{BEC}}$.

In general, the local single-site problem of a bosonic impurity coupled to
two baths (the condensate and normal bosons) has to be solved numerically.
The development of a reliable bosonic impurity solver \cite{bulla07} is a
challenging task, which took several years in the case of the fermionic
DMFT. This process can also involve a formulation of the proper bosonic
impurity Hamiltonian corresponding to the B-DMFT action derived here. One of
the main goals of this paper is to present the foundations of a novel,
comprehensive mean-field theory for correlated bosons and thereby instigate
further research by analytical and numerical means.

\begin{acknowledgments}
We thank R.~Bulla, W.~Hofstetter, and M.~Kollar for useful discussions. One
of us (DV) is grateful to V.~Janis for illuminating discussions during
1991-1992 on the scaling of the hopping amplitude of lattice bosons. This
work was supported in part by the Sonderforschungsbereich 484 of the
Deutsche Forschungsgemeinschaft (DFG).
\end{acknowledgments}

\appendix

\section{Derivation of the B-DMFT self-consistency equations}

Here we derive the B-DMFT equations for the generalized bosonic
Hubbard Hamiltonian (\ref{hamiltonian}) by applying the cavity
method. \cite{georges96} To this end the partition function
\begin{equation}
Z=\int D[b^*_{\nu},b_{\nu}]\exp(-S[b^*_{\nu},b_{\nu}])
\end{equation}
is calculated within the grand canonical ensemble, making use of the path
integral approach over complex coherent states.\cite{Negele} The action
\begin{eqnarray}
S[b^*_{\nu},b_{\nu}]&=&\int_0^{\beta}d\tau [\sum_{i\nu}
b^*_{i\nu}(\tau)(\partial_{\tau}-\mu)b_{i\nu}(\tau) + H(\tau)]
\end{eqnarray}
is split into a single-site term with $i=0$
\begin{eqnarray}
S_0=\int_0^{\beta}d\tau [b^*_{0\nu}(\tau)(\partial_{\tau}-\mu)b_{0\nu}(\tau)
+ \frac{1}{2}\sum_{\mu\nu}U_{\mu\nu} n_{0\mu} (n_{0\nu}-\delta_{\mu\nu})],
\end{eqnarray}
a term representing the coupling between this site and the rest of the
lattice ($i\neq 0$)
\begin{eqnarray}
\Delta S& =& \int_0^{\beta}d\tau \sum_{i\nu}\left(
t_{0i}^{\nu}b^{\dagger}_{0\nu} b_{i\nu}+ t_{i0}^{\nu}b^{\dagger}_{i\nu}
b_{0\nu}\right) \equiv \int_0^{\beta}d\tau \Delta S(\tau),
\end{eqnarray}
and a remaining part with site indices $i,j\neq 0$
\begin{eqnarray}
S^{(0)}=\int_0^{\beta}d\tau[ \sum_{i\neq 0\nu}
b^*_{i\nu}(\tau)(\partial_{\tau}-\mu)b_{i\nu}(\tau) + H^{(0)}(\tau)],
\end{eqnarray}
such that
\begin{eqnarray}
S[b^*_{\nu},b_{\nu}]= S_0+\Delta S+ S^{(0)}.
\end{eqnarray}
In the next step we expand the exponential function with respect to the
action $\Delta S$, and perform the functional integral over all variables
with site indices $i\neq 0$. As a result we obtain a formally infinite
series with all possible many-particle correlation functions, i.e.,
\begin{eqnarray}
Z=\int D[b^*_{0\nu},b_{0\nu}]e^{-S_0[b^*_{0\nu},b_{0\nu}]}Z^{(0)} \left(1-
\int_0^{\beta}d\tau \langle \Delta S(\tau)\rangle_{S^{(0)}} + \frac{1}{2!}
\int_0^{\beta}d\tau_1 \int_0^{\beta}d\tau_2 \langle \Delta S(\tau_1)\Delta
S(\tau_2) \rangle_{S^{(0)}}+ \cdot \cdot \cdot \right) ,
\end{eqnarray}
where $\langle ... \rangle_{S^{(0)}}$ denotes the average taken with
respect to $S^{(0)}$ (the action where the site $i=0$ is excluded)
and $Z^{(0)}$ is the corresponding partition function. In contrast
to the fermionic case \cite{georges96} there remain anomalous
correlation functions in the Bose-condensed phase such as $\langle
b_{i\nu} (\tau) \rangle_{S^{(0)}}$, $\langle b_{i\nu} (\tau)
b_{j\nu} (\tau^{\prime})\rangle_{S^{(0)}}$, or $\langle b_{i\nu}^*
(\tau) b_{j\nu} (\tau^{\prime})b_{k\nu}
(\tau^{\prime\prime})\rangle_{S^{(0)}}$, etc.
The lowest first-order terms take the form
\begin{eqnarray}
\int_0^{\beta}d\tau \langle \Delta
S(\tau)\rangle_{S^{(0)}}=\int_0^{\beta}d\tau
{\sum_{\nu}}{\sum_{j}}^{\prime}\left[ t^{\nu}_{0j} b_{0\nu}^*(\tau)
\langle b_{\nu j}(\tau) \rangle_{S^{(0)}} + t^{\nu}_{j0} b_{0\nu }
(\tau) \langle b_{j \nu }^*(\tau) \rangle_{S^{(0)}} \right],
\label{Z_1}
\end{eqnarray}
where the prime on the summation symbol indicates that the lattice indices
are different from $0$, i.e. $j\neq 0$ in (\ref{Z_1}). The second-order
terms read
\begin{eqnarray}
\frac{1}{2!} \int_0^{\beta}d\tau_1 \int_0^{\beta}d\tau_2 \langle
\Delta S(\tau_1)\Delta S(\tau_2) \rangle_{S^{(0)}} =\frac{1}{2!}
\int_0^{\beta}d\tau_1 \int_0^{\beta}d\tau_2 {\sum_{\nu}}{\sum_{jk}}^{\prime}  \nonumber \\
\left[ t_{j0}^{\nu} t_{k0}^\nu \langle b^*_{j\nu}(\tau_1) b^*_{k\nu}(\tau_2)
\rangle_{S^{(0)}} b_{0\nu}(\tau_1) b_{0\nu}(\tau_2) \right. +t_{j0}^{\nu}
t_{0k}^\nu \langle b^*_{j\nu}(\tau_1) b_{k\nu}(\tau_2) \rangle_{S^{(0)}}
b_{0\nu}(\tau_1) b_{0\nu}^*(\tau_2)  \nonumber \\
+t_{0j}^{\nu} t_{k0}^\nu \langle b_{j\nu}(\tau_1) b^*_{k\nu}(\tau_2)
\rangle_{S^{(0)}} b_{0\nu}^*(\tau_1) b_{0\nu}(\tau_2) +t_{0j}^{\nu}
t_{0k}^\nu \langle b_{j\nu}(\tau_1) b_{k\nu}(\tau_2) \rangle_{S^{(0)}}
b_{0\nu}^*(\tau_1) b_{0\nu}^*(\tau_2) \left. \right].  \label{Z_2}
\end{eqnarray}
Higher-order terms are obtained similarly. Defining the (connected)
correlation functions for the condensate
\begin{eqnarray}
\phi_{j\nu} (\tau)=\langle b_{j\nu}(\tau)\rangle_{S^{(0)}} \\
\phi_{j\nu}^* (\tau)=\langle b_{j\nu}^*(\tau)\rangle_{S^{(0)}},
\end{eqnarray}
(\ref{Z_1}) can be written as
\begin{eqnarray}
\int_0^{\beta}d\tau \langle \Delta
S(\tau)\rangle_{S^{(0)}}=\int_0^{\beta}d\tau
{\sum_{\nu}}{\sum_{j}}^{\prime}\left[ t^{\nu}_{0j} b_{0\nu}^*(\tau)
\phi_{j\nu}(\tau) + t^{\nu}_{j0} b_{0\nu } (\tau) \phi_{\nu
j}^*(\tau) \right].
\end{eqnarray}
Similarly, we define connected correlation functions for the one-particle
excitations above the condensate as
\begin{eqnarray}
G_{jk\nu}^{11\;(0)}(\tau_1-\tau_2)= - \langle T_{\tau}
b_{j\nu}(\tau_1)b_{k\nu}^*(\tau_2)\rangle_{S^{(0)}}, \\
G_{jk\nu}^{22\;(0)}(\tau_1-\tau_2)= - \langle T_{\tau}
b_{j\nu}^*(\tau_1)b_{k\nu}(\tau_2)\rangle_{S^{(0)}}, \\
G_{jk\nu}^{12\;(0)}(\tau_1-\tau_2)= - \langle T_{\tau}
b_{j\nu}(\tau_1)b_{k\nu}(\tau_2)\rangle_{S^{(0)}} , \\
G_{jk\nu}^{21\;(0)}(\tau_1-\tau_2)= - \langle T_{\tau}
b_{j\nu}^*(\tau_1)b_{k\nu}^*(\tau_2)\rangle_{S^{(0)}}.
\end{eqnarray}
which permits us to express the second-order contribution,
(\ref{Z_2}), as
\begin{eqnarray}
\frac{1}{2!} \int_0^{\beta}d\tau_1 \int_0^{\beta}d\tau_2 \langle \Delta
S(\tau_1)\Delta S(\tau_2) \rangle_{S^{(0)}} =- \frac{1}{2!}
\int_0^{\beta}d\tau_1 \int_0^{\beta}d\tau_2 {\sum_{\nu}}{\sum_{jk}}^{\prime}
\nonumber \\
\left[ t_{j0}^{\nu} t_{k0}^\nu G_{jk\nu}^{21\;(0)}(\tau_1-\tau_2)
b_{0\nu}(\tau_1) b_{0\nu}(\tau_2) \right. +t_{j0}^{\nu} t_{0k}^\nu
G_{jk\nu}^{22\;(0)}(\tau_1-\tau_2) b_{0\nu}(\tau_1)
b_{0\nu}^*(\tau_2)
\nonumber \\
+t_{0j}^{\nu} t_{k0}^\nu G_{jk\nu}^{11\;(0)}(\tau_1-\tau_2)
b_{0\nu}^*(\tau_1) b_{0\nu}(\tau_2) +t_{0j}^{\nu} t_{0k}^{\nu}
G_{jk\nu}^{12\;(0)}(\tau_1-\tau_2) b_{0\nu}^*(\tau_1)
b_{0\nu}^*(\tau_2)
\nonumber \\
-t_{j0}^{\nu} t_{k0}^\nu \phi_{j\nu}^*(\tau_1) \phi_{k\nu}^*(\tau_2)
b_{0\nu}(\tau_1) b_{0\nu}(\tau_2) -t_{j0}^{\nu} t_{0k}^{\nu}
\phi_{j\nu}^*(\tau_1) \phi_{k\nu}(\tau_2) b_{0\nu}(\tau_1)
b_{0\nu}^*(\tau_2)  \nonumber \\
-t_{0j}^{\nu} t_{k0}^\nu \phi_{j\nu}(\tau_1) \phi_{k\nu}^*(\tau_2)
b_{0\nu}^*(\tau_1) b_{0\nu}(\tau_2) -t_{0j}^{\nu} t_{0k}^{\nu}
\phi_{j\nu}(\tau_1) \phi_{k\nu}(\tau_2) b_{0\nu}^*(\tau_1)
b_{0\nu}^*(\tau_2) \left. \right].
\end{eqnarray}
Here the first four terms are due to connected contributions and the last
four terms due to disconnected contributions; higher terms can be written in
a similar way.

A non-trivial limit $d \rightarrow \infty$ is obtained by scaling
the hopping amplitudes $t^{\nu}_{ij}$ of $\nu$-bosons as described
in Sec. II.B. Namely, integer scaling is applied if $t_{ij}^{\nu}$
appears together with at least one anomalous average
$\phi_{i\nu}(\tau)=\langle b_{i\nu} (\tau)\rangle_{S^{(0)}}$
involving the BEC, while fractional scaling is employed otherwise.
For example, in the first-order term, (\ref{Z_1}), the sum over $j$
gives a contribution of the order $O(Z^{R_{0j}})$ so that the
hopping amplitude $t_{0j}^\nu$ must be scaled with (i.e. divided by)
a factor $Z^{R_{0j}}$ because $\phi_{j\nu}$ does not depend on the
distance. On the other hand, in the first four terms of the
second-order contribution to the partition function, (\ref{Z_2}),
the hopping amplitudes must be scaled with $Z^{R_{0j}/2}$ because
the one-particle correlation functions are already proportional to
$1/Z^{R_{0j}/2}$ as discussed in the Sec. II.B. In the last four
terms of the second order contribution the hopping amplitudes must
be scaled with $Z^{R_{0j}}$. In the calculation of higher-order
terms one has to distinguish the cases where all site indices are
different from those where some, or all, are the same. In analogy to
the fermionic case, discussed in detail in
Ref.~\onlinecite{georges96}, we find that all connected higher-order
terms vanish at least as $O(1/Z)$. Consequently, in the
$Z\rightarrow \infty$ limit only connected contributions containing
$\phi_{j\nu}$ or $G_{jk\nu}^{ab \;(0)}$, or disconnected
contributions made of products of connected contributions remain,
provided the infinite series converges at least conditionally.
Finally, we assume that the system is homogeneous, i.e., that
$\phi_{i\nu}=\phi_{\nu}$ is site independent. Applying the linked
cluster theorem and collecting only connected contributions in the
exponential function one obtains the local action
\begin{eqnarray}
S_{\mathrm{loc}}=\int_0^{\beta}d\tau
b^*_{0\nu}(\tau)(\partial_{\tau}-\mu)b_{0\nu}(\tau) + \int_0^{\beta}d\tau
\sum_{\nu} \kappa_{\nu} \left[ b_{0\nu}^*(\tau) \phi_{\nu}(\tau) + b_{0\nu }
(\tau) \phi_{\nu }^*(\tau) \right] -\int_0^{\beta}d\tau_1
\int_0^{\beta}d\tau_2 {\sum_{\nu}}{\sum_{jk}}^{\prime}  \nonumber \\
\left[ \tilde{t}_{j0}^{\nu} \tilde{t}_{k0}^\nu
G_{jk\nu}^{21\;(0)}(\tau_1-\tau_2) b_{0\nu}(\tau_1)
b_{0\nu}(\tau_2)\right. +\tilde{t}_{j0}^{\nu} \tilde{t}_{0k}^\nu
G_{jk\nu}^{22\;(0)}(\tau_1-\tau_2)
b_{0\nu}(\tau_1) b_{0\nu}^*(\tau_2)  \nonumber \\
+\tilde{t}_{0j}^{\nu} \tilde{t}_{k0}^\nu
G_{jk\nu}^{11\;(0)}(\tau_1-\tau_2) b_{0\nu}^*(\tau_1)
b_{0\nu}(\tau_2) + \left. \tilde{t}_{0j}^{\nu} \tilde{t}_{0k}^{\nu}
G_{jk\nu}^{12\;(0)}(\tau_1-\tau_2) b_{0\nu}^*(\tau_1)
b_{0\nu}^*(\tau_2)\right]  \nonumber \\
+\frac{1}{2}\sum_{\mu\nu}U_{\mu\nu} n_{0\mu} (n_{0\nu}-\delta_{\mu\nu}),
\end{eqnarray}
where the numerical factor $\kappa_{\nu}= \sum_{i\neq
0}\tilde{t}^{\nu}_{i0}/Z^{R_{i0}} $ for $d\rightarrow \infty$
depends on the lattice structure.

To simplify notations we introduce the Nambu formalism
\cite{Rickayzen} by defining a spinor boson operators
$\mathbf{b}_{i\nu}=(b_{i\nu},b^{\dagger}_{i,\nu})$ and corresponding
complex variables in the path-integral representation. Thereby
anomalous averages for the condensate
\begin{equation}
\mathbf{\Phi}_{i\nu}(\tau)\equiv \langle
\mathbf{b}_{i\nu}(\tau)\rangle_{S^{(0)}}
\end{equation}
and connected propagators for normal bosons
\begin{equation}
\mathbf{G}^{(0) }_{ij\nu}(\tau-\tau^{\prime})\equiv -\langle T_{\tau}
\mathbf{b}_{i\nu} (\tau) \mathbf{b}_{j\nu}^{\dagger}(\tau^{\prime})
\rangle_{S^{(0)}}
\end{equation}
can be written in a compact vector or matrix form. Introducing the
hybridization matrix function
\begin{equation}
\mathbf{\Delta}_{\nu}(\tau-\tau^{\prime})= - {\sum_{ij}}^{\prime}\tilde{t}%
^{\nu }_{i0}\tilde{t}^{\nu}_{j0}
\mathbf{G}_{ij\nu}^{(0)}(\tau-\tau^{\prime}),
\end{equation}
and employing the free (``Weiss'') mean-field propagator $\mathcal{G}_{\nu}$
one can express the B-DMFT local action in the form of (\ref{dmft0}). Here
the site index $i=0$ is omitted for simplicity.

Finally, the lattice self-consistency condition (\ref{dmft2}) needs
to be derived. For this we apply the relation between the Green
function $\mathbf{G}^{(0) }_{ij\nu}(\tau-\tau^{\prime})$ where the
site $i=0$ is removed and the full lattice Green function, i.e.
\begin{equation}
\mathbf{G}^{(0) }_{ij\nu}=\mathbf{G}_{ij\nu}-\mathbf{G}_{i0\nu}
\mathbf{G}_{00\nu}^{-1} \mathbf{G}_{0j\nu},
\end{equation}
which holds for a general lattice.  In the B-DMFT self-consistency equations
(\ref{dmft1}-\ref{dmft3}) for a homogeneous system only the site index $i=0$
enters which is therefore dropped.

\section{Free bosons on the Bethe tree with infinite coordination number}

In this Appendix we employ the B-DMFT to study a single species
($\nu =1$) of non-interacting bosons on the Bethe lattice with
$Z=\infty$.\cite{Eckstein,Kollar} Although this problem is exactly
solvable by different methods\cite{Berg} it is instructive to see
how the B-DMFT works in detail in this case.

\subsection{Green function method}

In order to obtain the Matsubara Green function
\begin{eqnarray}
G_{ij}(\tau-\tau^{\prime})=-\langle T_{\tau} b_i(\tau)
b^{\dagger}_j(\tau^{\prime})\rangle
\end{eqnarray}
for non-interacting bosons described by the Hamiltonian
(\ref{nonhamiltonian}) we use the Bogoliubov transformation to
separate the operator $b_i$ into a normal (non-condensate) part
$\tilde{b}_i$ and the condensate wave function $\phi_i$ as

\begin{eqnarray}
b_i=\tilde{b}_i+\phi_i  \nonumber \\
b_i^{\dagger}=\tilde{b}_i^{\dagger}+\phi_i^*.  \label{Bogol}
\end{eqnarray}
Assuming the system to be homogeneous, $\phi_i=\phi$, this yields
\begin{equation}
G_{ij}(\tau-\tau^{\prime})=-|\phi|^2+\tilde{G}_{ij}(\tau-\tau^{\prime}),
\end{equation}
where $\tilde{G}_{ij}$ is the Green function of the normal bosons.
In the non-interacting case considered here the anomalous Green function is
absent. The density of particles is given by
\begin{equation}
n=-\lim_{\tau^{\prime}\rightarrow
\tau^+}\frac{1}{N_L}\sum_iG_{ii}(\tau-\tau^{\prime})=
|\phi|^2-\lim_{\tau^{\prime}\rightarrow
\tau^+}\frac{1}{N_L}\sum_i\tilde{G}_{ii}(\tau-\tau^{\prime}),
\end{equation}
where $N_L$ is the number of lattice sites.

The diagonal Green function of normal bosons is given by
\begin{equation}
\tilde{G}_{ii}(\tau-\tau^{\prime})=\frac{1}{\beta}
\sum_{n}e^{-i\omega_n(\tau-\tau^{\prime})} \tilde{G}_{ii}(\omega_n),
\end{equation}
where
\begin{equation}
\tilde{G}_{ii}(\omega _n)= \frac{1}{N_L}\sum_{\lambda}\frac{1}{i\omega
_n+\mu-\lambda}= \frac{1}{i\omega_n+\mu-\Delta(\omega _n)},
\end{equation}
and $\lambda$ are the exact energy eigenstates of the lattice Hamiltonian.
The recursion relation
\begin{equation}
\tilde{G}_{ii}(\omega _n)=\frac{1}{i\omega_n+\mu-\tilde{t}^2
\tilde{G}_{ii}(\omega _n)},  \label{recursion}
\end{equation}
which is exact for the Bethe lattice,\cite{Eckstein,Kollar} allows one to
express the hybridization function as
\begin{equation}
\Delta(\omega_n) = \tilde{t}^2 \tilde{G}_{ii}(\omega_n).
\end{equation}
Eq.~(\ref{recursion}) determines $\tilde{G}_{ii}$ as
\begin{equation}
\tilde{G}_{ii}(\omega_n)=\frac{i\omega_n+\mu-
\sqrt{(i\omega_n+\mu)^2-4\tilde{t}^2}}{2\tilde{t}^2}. \label{Delta}
\end{equation}
In particular, the equation for the particle density follows as
\begin{equation}
n=|\phi|^2 -\frac{2}{\beta}\sum_{n} \frac{e^{i\omega_n0^+}}{i\omega_n+\mu+
\sqrt{(i\omega_n+\mu)^2-4\tilde{t}^2}}.  \label{G_tilde}
\end{equation}
Cauchy's theorem allows one to express the infinite sum as an integral over
the spectral function multiplied by the Bose-Einstein distribution function%
\cite{agd,Negele,Rickayzen} such that the density equation takes the form
\begin{equation}
n=|\phi|^2+\frac{1}{2\pi \tilde{t}^2}\int_{-2\tilde{t}}^{-2\tilde{t}}
d\omega \frac{\sqrt{4\tilde{t}^2-\omega^2}}{e^{\beta(\omega-\mu)}-1}.
\label{dens}
\end{equation}
For temperatures $T>T_{\mathrm{BEC}}$ the condensate vanishes,
$|\phi|=0$, in which case the equation determines $\mu$ as a
function of the density $n$. For $T<T_{\mathrm{BEC}}$ the chemical
potential is pinned at the value $\mu=-2\tilde{t}$ and (\ref{dens})
determines $|\phi|^2$, the density of the condensate. The
condensation temperature $T_{\mathrm{BEC}}$ itself is thus obtained
for $\mu=-2\tilde{t}$ and $|\phi|=0$. Expanding the Bose-Einstein
function into a Taylor series and changing the integration variable
into $\omega=2\tilde{t}\cos \theta$ one obtains a transcendental
equation for $T_{\mathrm{BEC}}$
\begin{equation}
n=\frac{4 \pi \tilde{t}^2 }{x}\sum_{k=1}^{\infty} \frac{e^{-kx}}{k}I_1(kx),
\end{equation}
where $x=2\tilde{t}/T_{\mathrm{BEC}}$ and $I_1$ is a modified Bessel
function.\cite{Abramowitz}

\subsection{B-DMFT}

We now show that the same results can be derived directly from the B-DMFT
equations. The local action takes the explicit form
\begin{eqnarray}
S_{\mathrm{loc}}=-\int_0^{\beta} d\tau \int_0^{\beta} d\tau
^{\prime}b^*(\tau)\mathcal{G}^{-1}(\tau - \tau^{\prime})b(\tau ^{\prime}) +
\tilde{t} \int_0^{\beta} d\tau \left( b^*(\tau) \phi(\tau) + \phi^*(\tau)
b(\tau) \right),
\end{eqnarray}
where the local Weiss Green function (an operator) is given by
\begin{equation}
\mathcal{G}^{-1}(\tau-\tau^{\prime}) = \delta(\tau-\tau^{\prime})\left( -
\partial_{\tau} + \mu \right) - \Delta(\tau-\tau^{\prime}).
\end{equation}
The hybridization function $\Delta(\tau)$ is determined
self-consistently by eqs.~(\ref{Dyson_eq},\ref{dmft2}). As in the
fermionic case the relation $\Delta(\omega_n) = \tilde{t}^2
\mathcal{G}(\omega_n)$ also holds for non-interacting bosons on the
Bethe lattice in the limit $Z\rightarrow \infty $.

In the absence of interactions the Euler-Lagrange equation of motion for the
classical field $\phi(\tau)$ is given by
\begin{eqnarray}
0=\frac{\delta S_{\mathrm{loc}}[b,b^*]}{\delta b^*(\tau)}\Bigg|%
_{b(\tau)=\phi(\tau)} =\;\;\; \left(\partial_{\tau} -\mu \right)\phi(\tau) +
\int_0^{\beta}d\tau^{\prime}\Delta (\tau - \tau^{\prime})\phi(\tau^{\prime})
+ \tilde{t} \phi(\tau).  \label{Euler}
\end{eqnarray}
By Fourier transformation (\ref{Euler}) becomes a linear equation
\begin{equation}
\left( i\omega_n +\mu-\tilde{t}-\Delta (\omega_n)\right) \phi(\omega_n)=0.
\label{linear}
\end{equation}
Employing (\ref{Delta},\ref{G_tilde}) 
this equation takes the form
\begin{equation}
\left(i\omega_n+\mu - 2\tilde{t} + \sqrt{(i\omega_n+\mu)^2-(2\tilde{t})^2}
\right)\phi(\omega_n) =0.  \label{phi}
\end{equation}
In the static limit $\omega_n=0$, corresponding to $n=0$, this
equation has the solution $\phi=0$ when $\mu <- 2\tilde{t}$,
implying that the chemical potential lies outside the bosonic band,
or the solution $\phi\neq 0$ when $\mu= - 2\tilde{t}$. In the latter
case the actual value of $|\phi|$ (which determines the BEC
fraction) must be computed from the equation for the particle
density, (\ref{dens}). In the dynamical case, $\omega_n\neq 0$, i.e.
for $n\neq 0$, (\ref{phi}) only has the solution $\phi(\omega_n)=0$
because the expression in the bracket never vanishes. This shows
that for non-interacting bosons the condensate order parameter is
time independent.

After Fourier transformation the local action takes the form
\begin{equation}
S_{\mathrm{loc}} = \sum_{n} b_n^* \left[ i\omega_n+\mu-
\Delta(\omega_n)\right]b_n +\tilde{t}\left(b_{n=0}^*\phi +
b_{n=0}\phi^* \right),
\end{equation}
where the numbers $b_n$ are the Fourier coefficients of $b(\tau)$ in
Matsubara frequency space. The zero frequency component $b_{n=0}$ is
shifted by the Bogoliubov transformation (\ref{Bogol}) as
$b_{n=0}=\tilde{b}_{n=0}+ \phi $. Because of (\ref{linear}) the
local action is seen to be quadratic in $\tilde{b}_n$, and the
functional integral yields the same equation for the particle
density as (\ref{dens}). Thus we showed that the B-DMFT correctly
reproduces all results for non-interacting bosons on the Bethe
lattice both in the normal and in the BEC phase.


\end{document}